\documentstyle[twoside,fleqn,espcrc2,fps]{article}
 
\newcommand{\chibar}{\overline{\chi}}
\newcommand{\Z}{Z\!\!\!Z}

\newcommand{\un}{1\!\!1}
\newcommand{\half}{\frac{1}{2}}

\newcommand{\dsl}{\partial \!\!\!/}

\newcommand{\be}{\begin{equation}}
\newcommand{\ee}{\end{equation}}
 
\def \3{\ss}

\title{Perfect Lattice Actions for the Gross--Neveu Model}
 
\author{Wolfgang Bietenholz\address{Center for Theoretical Physics,
            Massachusetts Institute of Technology(MIT)\\
            Cambridge, Massachusetts 02139, USA},
        Erich Focht\address{Institute for Theoretical Physics E,
            RWTH-Aachen, 52074 Aachen, Germany\\
            HLRZ c/o KFA J\"ulich, 52425 J\"ulich, Germany}
        and
        Uwe-Jens Wiese$^{\rm a}$
       \thanks{Based on two talks presented by E.F. and W.B.
               Work supported by Deutsche Forschungsgemeinschaft
(D.F.G.) and U.S. Department of Energy (D.O.E.) under cooperative
research agreement DE-FC02-94ER40818.}
}
\begin{document}
 
\begin{abstract}
We apply the method of Hasenfratz and Niedermayer to analytically
construct perfect lattice actions for the Gross--Neveu model. In the
large $N$ limit these actions display an exactly perfect scaling,
i.e. cut-off artifacts are completely eliminated
even at arbitrarily short correlation length. Also the energy
spectrum coincides with the spectrum in the continuum
and continuous translation and rotation symmetries are restored
in physical observables.
This is the first (analytic) construction of an exactly
perfect lattice action at finite correlation length.
\end{abstract}
 
\maketitle
 
 
\section{Introduction}
 
The worst systematic errors in lattice simulations are due to
cut-off artifacts. For fermionic systems they are typically
of $O(a)$, where $a$ is the lattice spacing. For a long time,
the attempts to improve the lattice action such that the
artifacts are dampened, yielded only a limited success.
Last year, P. Hasenfratz and F. Niedermayer proposed a new
method for approximating perfect actions (actions free of
cut-off effects) for asymptotically free theories \cite{HN}.
It might have the potential for a break-through which enables
us to solve lattice QCD.
 
First we note that the fixed point actions (FPAs)
- obtained from
iterated block spin renormalization group transformations (RGTs)
on a critical surface
(see e.g. \cite{Ma}) - are perfect actions. However, since the correlation
length $\xi$ diverges there, they cannot be used in simulations.
Kogut and Wilson \cite{KW} noticed that perfect actions exist also
off the critical surface. They lie on `renormalized trajectories'
(RTs) in parameter space, which cross the critical surface
in a fixed point. The direction of the tangent there corresponds
to a relevant (or marginal) direction. If we move slightly away
from a fixed point in a relevant direction and perform RGTs, then we
follow the RT. Thus all points on it are related to the vicinity
of the fixed point without occurrence of irrelevant terms.
 
For asymptotically free theories, there is only one (weakly)
relevant (to leading order marginal) direction. Hasenfratz
and Niedermayer propose to follow this direction down to a
$\xi$ short enough for simulation, i.e. to approximate the RT
linearly. The points on this tangent are referred to as
`perfect classical actions' and there is hope that they
still have very small cut-off artifacts.
The virtue of asymptotically free theories (in this context) is
that their FPAs can be determined classically. 

The pilot project of Hasenfratz and Niedermayer dealt with the
$O(3)$ nonlinear $\sigma$ model in $d=2$ and was a full success.
Even on a lattice as small as $5 \times 5$ cut-off effects
were numerically not visible any more.
 
The final goal is of course QCD. A collaboration is presently
studying the pure $SU(3)$ gauge theory in this respect \cite{AH}.
 
Here we present results for the 2d Gross--Neveu model \cite{BFW}, which
turned out to be a good testing ground for the method when applied
to fermionic systems.
 
\section{Fixed point actions for free lattice fermions}
 
First, one of us found a line of FPAs for free Wilson fermions
\cite{UJW} which also applies to a number of other lattice
fermions (SLAC, Rebbi, etc., see \cite{WB}). However, this set does not
include the staggered fermions which we are going to use here.
Their main advantage is the remnant $U(1) \otimes U(1)$ symmetry
which persists at free, local fixed points \cite{Dal}.\footnote{For
Wilson fermions the full chiral symmetry may be preserved,
but then the FPA becomes non local, in agreement with the
Nielsen-Ninomiya theorem.}
We perform RGTs according to the prescription in \cite{KMS} i.e.
we attach pseudo flavors to the corners of (disjoint) $2 \times 2$
blocks and build
block spins such that the pseudo flavors don't mix. Therefore
an odd blocking factor is needed; we chose it to be 3. Then we
have intersecting $9 \times 9$ blocks, whose centers form
a lattice of spacing 3 and which are occupied by alternating pseudo
flavors. Every point in a block with the same pseudo flavor as the
center contributes to its block spin which lives again on the block
center. Then we rescale coordinates and fields.
 
We take a lattice of spacing $1/2$. For technical reasons we
initially put the fermionic pseudo flavors $\chibar^{1}, \dots
\chibar^{4}, \chi^{1} \dots \chi^{4}$ (one--component Grassmann
variables) on the centers $x$ of disjoint $2 \times 2$ blocks
(spacing 1). Later the fermionic variables
are moved to the corners (in \cite{Dal} we proceeded differently).
For the action after a number of RGTs we make the general ansatz
\begin{equation}
S[\chibar, \chi ] = \sum_{x,y} \sum_{i,j=1}^{4}
\chibar^{i}_{x} \rho^{ij}(x - y ) \chi^{j}_{y} \ .
\end{equation}
$S$ is invariant against translations by integers (but not half integers)
along the coordinate axes.
The matrix for the fermions shifted to the corners is (in
momentum space) denoted by  $\tilde \rho (k)$. Its form is restricted
by charge conjugation, $U(1)\otimes U(1)$ symmetry and half integer
translations. Its structure, resp.
the one of its inverse $\tilde \alpha (k)
= \tilde \rho^{-1}(k)$ is
\begin{equation} \label{alpha}
\tilde \alpha (k) \! = \! \left( \! \begin{array}{cccc} 0 & 
\tilde \alpha_{1}(k) &
\tilde \alpha_{2}(k) & 0 \\ \tilde \alpha_{1}(k) & 0 & 0 & -
\tilde \alpha_{2}(k) \\ \tilde \alpha_{2}(k) & 0 & 0 &
\tilde \alpha_{1}(k) \\ 0 & -\tilde \alpha_{2}(k) &
\tilde \alpha_{1}(k) & 0 \end{array} \! \right) .
\end{equation}
Starting from the standard action with
$\tilde \alpha_{\mu}(k) = i \hat k_{\mu}
/ \hat k ^{2} $ (in lattice notation: $\hat k_{\mu} \doteq 2 \sin (k_{\mu}
/2) )$, we arrive at the fixed point given by
\begin{eqnarray}
  \tilde \alpha^*_\mu(k) & \! \! = & \! \! i\sum\limits_{l\in \Z^2}
      \frac{k_\mu+2\pi l_\mu}{(k+2\pi l)^2} (-)^{l_\mu}
      \prod\limits_\nu ( \frac{\hat k_\nu}{k_\nu+2\pi l_\nu})^2
      \nonumber \\
   & \! \! +& \! \! \frac{9}{8c}i \hat k_\mu.
\end{eqnarray}
Here a renormalization parameter in the blocking step had to
be chosen in agreement with the dimension of $\chi$. The case
$c \to \infty$ corresponds to a $\delta $--function blocking
RGT, but we may add a `smearing term' suppressed by $1/c$,
which does not break any symmetry and which helps to optimize
the locality, see \cite{Dal}.
 
 
\section{The Gross--Neveu model}
 
Our main interest was to gain experience with interacting fermionic
models, the simplest asymptotically free one being the Gross--Neveu
model in two dimensions (GN). It is a model with 4-Fermi interaction
which can be linearized by introducing a real valued auxiliary field
$\Phi$. With $G$ being the usual 4-Fermi coupling the continuum
Euclidean action for $N$ flavors is
\begin{displaymath}
\frac{1}{G}S[\chibar,\chi,\Phi]  \! = \! \frac{1}{G}\!
  \int\!\!\! d^{2}x \! \! \sum\limits_{i=1}^N (\chibar^i \dsl \chi^i
                                  +\chibar^i\chi^i\Phi)
                  +\half \Phi^2 .
\end{displaymath}
The $O(2N)\times Z(2)$ symmetry of the model is broken spontaneously
to $O(2N)$ and a fermion mass $m_f$ is generated dynamically.

In a sensible lattice formulation of the GN model with staggered
fermions and Yukawa coupling, the scalars live on the centers of the
fermionic lattice, i.e. they couple equally to the 4 pseudo flavors at
one plaquette and their spacing is 1/2. To keep a Gaussian form we do
not integrate out $\Phi$ but block it, simultaneous to the fermionic
blocking. For this we use the same pattern: we block them as if they
also had pseudo flavors, which must not be mixed, see fig.1.
 \begin{figure}[htb]
 \begin{center}
 \begin{minipage}{4.0 cm}{}
 \begin{picture}(115,115)(10.0,10.0)
 \multiput(10,10)(32,0){4}{$\bullet$}
 \multiput(26,10)(32,0){4}{$\circ$}
 \multiput(10,26)(32,0){4}{$\diamond$}
 \multiput(26,26)(32,0){4}{$\ast$}
 \multiput(10,42)(32,0){4}{$\bullet$}
 \multiput(26,42)(32,0){4}{$\circ$}
 \multiput(10,58)(32,0){4}{$\diamond$}
 \multiput(26,58)(32,0){4}{$\ast$}
 \multiput(10,74)(32,0){4}{$\bullet$}
 \multiput(26,74)(32,0){4}{$\circ$}
 \multiput(10,90)(32,0){4}{$\diamond$}
 \multiput(26,90)(32,0){4}{$\ast$}
 \multiput(10,106)(32,0){4}{$\bullet$}
 \multiput(26,106)(32,0){4}{$\circ$}
 \multiput(10,122)(32,0){4}{$\diamond$}
 \multiput(26,122)(32,0){4}{$\ast$}
 \multiput(17,18)(16,0){7}{$\times$}
 \multiput(17,34)(16,0){7}{$\times$}
 \multiput(34,34)(32,0){3}{$\circ$}
 \multiput(17,50)(16,0){7}{$\times$}
 \multiput(17,66)(16,0){7}{$\times$}
 \multiput(34,66)(32,0){3}{$\circ$}
 \multiput(17,82)(16,0){7}{$\times$}
 \multiput(17,98)(16,0){7}{$\times$}
 \multiput(34,98)(32,0){3}{$\circ$}
 \multiput(17,114)(16,0){7}{$\times$}
 \thicklines
 \put(32.5,32.00){\framebox(72,72){\mbox}}
 \multiput(20.8,20.5)(48,0){3}{\circle{6}}
 \multiput(20.8,68.5)(48,0){3}{\circle{6}}
 \multiput(20.8,116.5)(48,0){3}{\circle{6}}
 \end{picture}
 \end{minipage}
 \end{center}
 \vspace{-1cm}
 \caption{\it Geometry of the scalar (crosses) and fermionic variables
 (different symbols for the pseudo flavors).
 The nine auxiliary field variables marked with a small circle
 contribute to the block variable in the center of the box.} 
 \vspace{-0.5cm}
 \label{GNfigure}
 \end{figure}
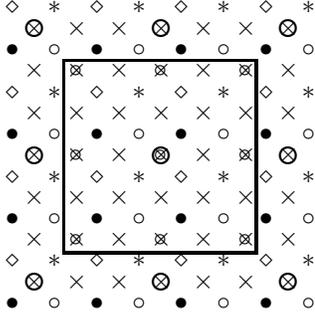

\section{Small field approximation}
 
For the moment we only include the first order in $\Phi $ and let $N=2$.
Then we parametrize a general ansatz for the action as:
\begin{eqnarray} \label{ANSATZ}
 \frac{1}{G}S[ \chibar , \chi ,\Phi ]
& \! \! =  & \! \! \frac{1}{G} \{ \sum\limits_{x,y}
                                 \sum\limits_{i,j}
                                \chibar_{x}^i \rho^{ij}(x-y)
                                \chi_{y}^j \\
                                + \half\sum\limits_z \Phi_z^2
    \!\!\!\!  & \! \!  \!  + & \! \! \! \! \! \!
                  \sum\limits_{x,y,z} \sum\limits_{i,j}
                  \chibar_{x}^i\sigma^{ij}(x-z,y-z)
                  \chi_{y}^j\Phi_z \}. \nonumber
\end{eqnarray}
where $z$ runs over the cell centers (spacing $1/2$).
For the standard action in momentum space, shifting the fermions
to the block corners takes $\sigma $ to the form
\begin{eqnarray}
\tilde \sigma (p,q) &=& \left(
  \begin{array}{cccc}
     \tilde \sigma_{0} & 0 & 0 & -\tilde \sigma_{3} \\
     0 & \tilde \sigma_{0} & \tilde \sigma_{3} & 0 \\
     0 & -\tilde \sigma_{3} & \tilde \sigma_{0} & 0 \\
     \tilde \sigma_{3} & 0 & 0 & \tilde \sigma_{0}
  \end{array}
 \right) \ , \\
\tilde\sigma_{0} &=& \frac{\hat p_\mu \hat q_\mu}{\hat p^2 \hat q^2}
               \prod_{\nu} \cos ((p_{\nu}+q_{\nu})/4) \ , \nonumber \\
\tilde\sigma_{3} &=&
               \frac{\hat p_1\hat q_2-\hat p_2\hat q_1}{\hat p^2\hat q^2}
               \prod_\nu \cos ((p_\nu + q_\nu)/4) \ .
               \nonumber
\end{eqnarray}
The fixed point condition forces us to apply a $\delta $ blocking
in the scalar sector. Doing so we arrive at a fixed point, where
$\sigma $ still has the same structure and its elements become:
\begin{eqnarray}
\tilde \sigma_{0}^{*} & \! \! = & \! \! \sum\limits_{l,m\in\Z^2}\sum_i
                   \frac{p^{(l,i)}_\mu q^{(m,i)}_\mu}{(p^{(l,i)})^2
                         (q^{(m,i)})^2} \label{sfa} \\
               & \! \! \times & \! \! \prod_\nu (-)^{l_\nu+m_\nu+i_\nu}
                       \frac{\hat p_\nu \hat q_\nu
                             \widehat{(p+q)_\nu} }{p^{(l,i)}_\nu
                             q^{(m,i)}_\nu
                             (p^{(l,i)}_\nu+q^{(m,i)}_\nu)} ,
                   \nonumber\\
\tilde \sigma_{3}^{*} & \! \! =& \! \! \sum\limits_{l,m\in\Z^2}\sum_i
                   \frac{\epsilon_{\mu\rho} p^{(l,i)}_\mu
                          q^{(m,i)}_\rho}{(p^{(l,i)})^2
                          (q^{(m,i)})^2}
                   \nonumber \\
               & \! \! \times & \! \! \prod_\nu (-)^{l_\nu+m_\nu}
                       \frac{\hat p_\nu \hat q_\nu
                             \widehat{(p+q)_\nu} }{p^{(l,i)}_\nu
                             q^{(m,i)}_\nu
                             (p^{(l,i)}_\nu+q^{(m,i)}_\nu)}
                   \nonumber ,
\end{eqnarray}
with $p^{(l,i)}_\mu=p_\mu+4\pi l_\mu+2\pi i_\mu$ and $i_{\mu}\in\{0,1\}$.
The $4\pi $ antiperiodicity of these elements agrees with the
central positions of the scalars.
Thus we have identified the direction of the RT out of the critical
surface, but only in the parameter subspace which is accessible
in the framework of the small field approximation.

\section{The large $N$ limit}
 
In general one can hardly go beyond the small field approximation
to determine the functional $S^{*}[\chibar , \chi , \Phi]$
analytically. An exception is the limit $N \to \infty $.
There the scalar field freezes to a constant field $\Phi_{0}$
and we can calculate $S^{*}[\chibar ,\chi , \Phi_{0}]$
nonperturbatively. We parametrize a general ansatz as
\begin{eqnarray}
&& S[\chibar,\chi,\Phi_0] = \\ && \! \! \! \! 
          \frac{1}{(2\pi )^{2}} \!
          \int_B \! d^{2}k \chibar(-k) [\rho (k) + \lambda (k) \un]
          \chi(k) + \frac{1}{2} \Phi_{0}^{2} V \nonumber
\end{eqnarray}
where $B$ is the Brillouin zone and $V$ is the space-time volume factor
(actually infinite). For the standard action $\rho $ is the
same as for free fermions and $\lambda (k) = \Phi_{0}$.
We define $\tilde \alpha (k) + \beta ( k)\un \doteq [\tilde \rho (k)
+ \lambda (k) \un]^{-1}$ (again the tilde corresponds to fermions
shifted to the plaquette corners; for $\lambda $ and $\beta $
there is no change). Still $\tilde \rho $ and $\tilde \alpha $
keep the structure given in (\ref{alpha}) and we obtain at the
fixed point
\begin{eqnarray} \label{exact}
\tilde \alpha^{*}_{\mu}(k) &\! \! =&\! \! i\sum_{l \in \Z^{2}} \frac{k_{\mu}+
   2\pi l_{\mu} }{(k+2\pi l)^{2}+\Phi_{0}^{2}} (-1)^{l_{\mu}} \\
&\! \! &\! \! \times \prod_{\nu} \Big( 
   \frac{ \hat k_{\nu}}{k_{\nu}+2\pi l_{\nu} }
   \Big) ^{2} + \frac{9}{8c} i \hat k_{\mu} \ ,
   \nonumber \\
\beta^{*}(k) &\! \! =&\! \! \sum_{l \in \Z^{2}} \!
   \frac{\Phi_{0}}{(k+2\pi )^{2}
   + \Phi_{0}^{2}} \prod_{\nu} \Big( \frac{ \hat k_{\nu}}{k_{\nu}
   +2\pi l_{\nu}} \Big)^{2} \! .
   \nonumber
\end{eqnarray}
Here the small field approximation allows to include small non--zero
modes of $\Phi (k)$ and may be used if one is interested e.g.
in $1/N$ effects. The result to the first order in $\Phi (k \neq 0)$
is given in \cite{BFW}. The occurring matrix $\tilde \sigma^{*}(p,q)$ is more
general than the one given in (\ref{sfa}), because we expand around
any $\Phi_{0}$ now, whereas (\ref{sfa}) describes a particular expansion
around $\Phi_{0} = 0$.

\section{The perfect operator for the chiral condensate}
 
Let's perturb $S^{*}$ a little with some operator $X$:
\begin{equation}
S_{j}[\chibar ,\chi , \Phi ] = S^{*}[\chibar , \chi , \Phi ]
+ j X[\chibar , \chi , \Phi ] \ ,
\end{equation}
where $j \ll 1$.
We call $X$ a {\em perfect operator} and denote it by $X^{*}$
if it is an eigenfunctional of the RGT to $O(j)$
\begin{eqnarray} \label{cpo}
S'_{j}[\chibar',\chi',\Phi'] & = & S^*[\chibar',\chi',\Phi'] \\
&& + j\gamma X^*[\chibar',\chi',\Phi'] + O(j^2) . \nonumber
\end{eqnarray}
The operator is relevant if $\gamma > 1$. This is the case for the
chiral condensate (with standard operator $\sum_{x} \chibar_{x}
\chi_{x}$): for dimensional reasons $\gamma $ coincides with the
block factor.
 
We specialize again to $N\to\infty$, $\Phi\to\Phi_0$ and also
parametrize $X$ as:
\begin{displaymath}
X[\chibar,\chi,\Phi_0] \! = \!\! \frac{1}{(2\pi)^2} \! \!\int_B \!
\!dk \chibar(-k) [\mu (k)+ \nu (k) \un] \chi(k) ,
\end{displaymath}
where $\mu$ has the structure of $\rho $ and $\alpha $.
From the above condition (\ref{cpo}) for $X^{*}$ we obtain
(again shifting the fermions to the block corners)
\begin{eqnarray} \label{po}
\tilde \mu^{*}_{\mu}(k) &\! \! = & \! \!
\partial_{\Phi_{0}} \frac{\tilde \alpha_{\mu}^{*}(k)}
{\tilde \alpha_{\nu}^{*}(-k)
\tilde \alpha_{\nu}^{*}(k) + \beta^{*}(-k) \beta^{*} (k)} , \\
\nu^{*}(k) & \! \! = & \! \! 
\partial_{\Phi_{0}} \frac{\beta^{*}(k)}{\tilde \alpha_{\nu}
^{*}(-k) \tilde \alpha_{\nu}^{*}(k) + \beta^{*}(-k) \beta^{*}(k)} . \nonumber
\end{eqnarray}
(Omitting $O(\Phi_{0})$ we observe that $X^{*}$ is very local.)

\section{The gap equation}
 
Now we forget about all small field approximations, let
$N \to \infty $ keeping $g = GN$ fixed and exploit our exact
result (\ref{exact}). We make use of the dimensional transmutation
due to the spontaneous symmetry breaking $U(N/2) \otimes Z(2)
\to U(N/2)$, which gives rise to a fermion mass $m_{f}$.
We introduce an effective potential $V_{eff}$ by
\begin{equation}
e^{-V_{eff}(\Phi_{0}) V} = \int D \chibar D \chi
e^{-\frac{1}{G} S [ \chibar , \chi , \Phi_{0}]} \ .
\end{equation}
The condition for $V_{eff}$ to be minimal yields the gap equation
\begin{equation}
2 \Phi_{0} = \frac{1}{(2\pi )^{2}} \int_{B} d^{2}k \ g \ln \det
{\cal M} (k,\Phi_{0}) \ ,
\end{equation}
${\cal M}=-\frac{1}{G} [\tilde\rho + \lambda\un] $ being the fermion
matrix for one set of staggered fermions. For the standard action
the gap equation takes the form
\begin{equation} \label{stagap}
\frac{1}{g} = \frac{2}{(2 \pi )^{2}} \int_{B} d^{2}k \frac{1}
{\hat k^{2}+\Phi_{0}^{2}} \ .
\end{equation}
obtained and discussed before in \cite{Bel}. For the FPA it reads
\begin{eqnarray}
\Phi_{0} &=& \frac{1}{(2\pi )^{2}} \int_{B} d^{2}k \ g \partial_{\Phi_{0}}
\ln [ \tilde \alpha_{\mu}^{*}(-k) \tilde \alpha_{\mu}^{*}(k) \nonumber \\
&& + \beta^{*}(-k) \beta^{*}(k)] \label{fpagap} \ .
\end{eqnarray}
First we investigate asymptotic scaling of $\Phi_{0}$, see fig.2.
Asymptotic scaling corresponds to linear curves.
We see that asymptotic scaling for the FPA is clearly
better, but still not what one would call `perfect'.
As an alternative consideration of asymptotic scaling we compare
in fig.3 the $\beta $
functions referring to the fermion mass, which we will derive below.
It shows the same feature. (Asymptotic
scaling is given by: $a m_{f} \propto e^{-\pi / g} \ \Rightarrow
\beta = - m_{f} \partial _{m_{f}} g = - g/\pi^{2}$.) At $\xi =1$
the FPA's $\beta $ function already agrees with this
up to 5 percent, but its behavior is still not perfect.

\begin{figure}[htb]
 \centerline{
  \def\fpsangle{90}
  \epsfxsize=6cm
  \fpsbox{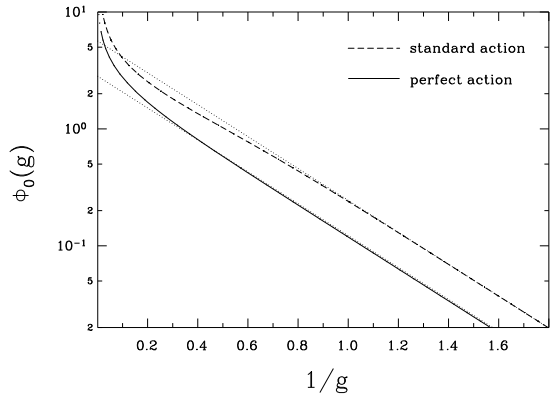}
 }
 \vspace{-1cm}
 \caption{\it Asymptotic scaling behavior of $\Phi_0(g)$.}
 \vspace{-0.5cm}
 \label{asympscaling}
\end{figure}
\begin{figure}[htb]
 \centerline{
  \def\fpsangle{90}
  \epsfxsize=6cm
  \fpsbox{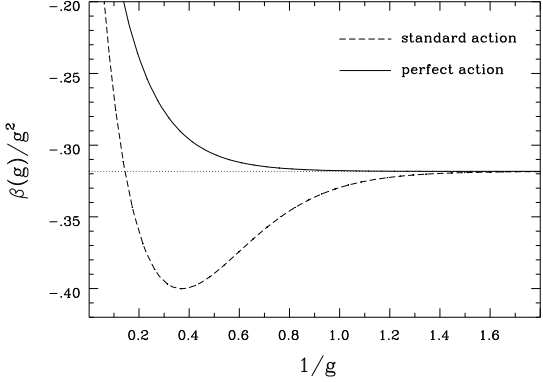}
 }
 \vspace{-1cm}
 \caption{\it $\beta(g)/g^2$ versus $1/g$.}
 \vspace{-0.5cm}
 \label{betafunction}
\end{figure}

However, this doesn't mean very much. Asymptotic scaling is not really
physical since it involves the bare coupling (at cut-off scale).
Actually meaningful are {\em dimensionless ratios}.
 
\section{The scaling behavior}
 
Such a ratio shall be constructed now to discuss the scaling
behavior of the standard action, the small field approximation,
and finally of the FPA. First we derive $m_{f}$
from the exponential decay of the correlation function.
We put the system in a spatial volume $L$ and consider $x_{2}$
as Euclidean time. The correlation function becomes
\begin{eqnarray}
&& \langle \chibar (-k_{1})_{0} \chi (k_{1})_{x_{2}} \rangle = \nonumber \\
&& \frac{LN}{4 \pi}
\int_{-\pi}^{\pi} dk_{2} \ {\rm Tr} \ {\cal M}^{-1}(k,\Phi_{0}) 
e^{ik_{2}x_{2}} \ .
\end{eqnarray}
For the standard action this turns into
\begin{displaymath}
- \frac{L}{\pi} g \Phi_{0} \int_{-\pi}^{\pi} dk_{2} \frac{
e^{ik_{2}x_{2}}}
{\hat k^{2} + \Phi_{0}^{2}} \doteq C(k_{1}) e^{-E(k_{1})x_{2}} .
\end{displaymath}
At the pole of the integrand we find the energy spectrum:
\begin{equation}
[ 2 \sinh ( E(k_{1})/2) ]^{2} = \hat k_{1}^{2} + \Phi_{0}^{2}
\end{equation}
We observe cut-off artifacts of $O(a^{2})$. We identify
$m_{f} = E(0)$ such that
\begin{displaymath}
2 \sinh (m_{f}/2) = \Phi_{0} \ .
\end{displaymath}
 
Now we consider the chiral condensate
\begin{equation}
\langle \chibar \chi \rangle = \frac{1}{(2\pi )^{2}} \frac{N}{2} \int_{B}
d^{2} k \ {\rm Tr} \ {\cal M}^{-1}(k, \Phi_{0}) .
\end{equation}
For the standard action the gap equation (\ref{stagap}) leads to
$\langle \chibar \chi \rangle = - \Phi_{0}$. Therefore
\begin{equation}
\frac{\langle \chibar \chi \rangle }{m_{f}} \vert_{{\rm standard}} =
- \frac{ 2 \sinh (m_{f}/2)}{m_{f}} \ .
\end{equation}
In the continuum limit, this ratio must become constant, and indeed
it does, since: $\xi = m_{f}^{-1} \to \infty $ (in lattice units).
For finite $\xi $, however, we have artifacts of $O(a^{2})$.
The $Z(2)$ symmetry shields the system from $O(a)$ artifacts.
 
We don't go through the corresponding formulae for the small field
approximation here but just quote the result that the scaling
artifacts remain of $O(a^{2})$ \cite{BFW}. There is a slight improvement
on the $O(a^{2})$ level, but this is not a great progress.
 
Finally we consider the dispersion relation for the FPA:
\begin{eqnarray}
&& \! \! \! \! \langle \chibar (-k_{1})_{0} \chi (k_{1})_{x_{2}}
\rangle \ =  \\
&&\! \! \! \! - \frac{L}{\pi} g \Phi_{0}
\int_{-\pi}^{\pi} dk_{2} \{ \sum\limits_{l \in \Z^{2}} \frac{e^{ik_{2}x_{2}}}
{(k+2\pi l )^{2} + \Phi_{0}^{2}} \nonumber \\ && \! \! \! \!
\prod_{\nu} \Big( \frac{ \hat k_{\nu}}
{k_{\nu}+2\pi l_{\nu}} \Big)^{2} \} \nonumber \\
&\! \! = & \! \! \! \! -\frac{Lg\Phi_{0}}{\pi}\! \int_{-\infty}^{\infty}\! \!
dk_{2} \{ \sum\limits_{l_{1} \in \Z } \frac{e^{ik_{2}x_{2}}}
{(k_{1}+2\pi l_{1})^{2} + k_{2}^{2}
+ \Phi^{2}_{0} } \nonumber \\ && \! \! \! \!
\Big( \frac{\hat k_{1}}{k_{1} +2\pi l_{1}} \cdot
\frac{\hat k_{2}}{k_{2}}
\Big) ^{2} \} \nonumber \\
&\! \! \doteq & \! \!
\sum\limits_{l_{1} \in \Z} C(k_{1} + 2\pi l_{1}) \exp (-E
(k_{1}+2\pi l_{1})x_{2}) \nonumber
\end{eqnarray}
For each $l_{1}$ a pole contributes to the correlation function
and hence
\begin{equation}
E^{2} ( k_{1}+2\pi l_{1}) = - k_{2}^{2} = (k_{1} + 2\pi l_{1})^{2}
+ \Phi_{0}^{2} \ .
\end{equation}
This is the {\em exact} continuum energy spectrum. It is remarkable
that it is obtained although $k_{1} \in ]-\pi , \pi ]$. Here we
also observe the restoration of continuous rotational symmetry
resp. Lorentz invariance (which can not be seen directly in
$\tilde \alpha (k)$.) Moreover also the continuous translation
invariance is restored (note that $k_{1}+2\pi l_{1}$ covers
the whole real axis).
 
Using the perfect operator for the chiral condensate (\ref{po}) and
inserting the FPA's gap eq. (\ref{fpagap}), we obtain also for the FPA
\footnote{Both equations are quite complicated, but they have
exactly the same structure, which allows for a drastic
simplification.}
\begin{equation}
\langle \chibar \chi \rangle = - \Phi_{0} \ .
\end{equation}
Now the dimensionless ratio becomes
\begin{equation}
\frac{\langle \chibar \chi \rangle }{m_{f}} \vert_{{\rm FPA}} = -1 \ ,
\end{equation}
for any $\xi$, i.e. we have {\em perfect scaling}.
 
The result that in the FPA cut-off effects are completely
eliminated means that the perfect classical actions are situated
on the renormalized trajectory. This can be understood
from the fact that in the limit $N \to \infty $ the path integral
reduces to the classical solution; all fluctuations around
the minimum are suppressed with weight zero.
 
\section{Summary and outlook}
 
What we should know to determine the classically perfect action
for any $N$ is
the FPA $\frac{1}{G} S^{*}[\chibar ,\chi ,\Phi ]$, which
is reproduced under RGT. However, this could be
obtained analytically only in the small $\Phi $ approximation.
 
If we let $N \to \infty $, it suffices to consider a constant
$\Phi_{0}$ and $S^{*}$ can be calculated exactly (nonperturbatively).
Then the large $N$ limit does a second job for us and makes the
line of perfect classical actions coincide with the renormalized trajectory,
such that we indeed observe perfect scaling.
 
For finite $N$, both trajectories will change (there occur
higher order interactions etc.) and we have no reason to expect
that they still coincide. It would be interesting to determine
the size of the cut-off artifacts on the classical trajectory
in this case. This could be done either numerically for
finite $N$, similar to \cite{HN}, or by a $1/N$ expansion.
A basis for the latter is given by our results for the small
field approximation applied to the non--zero modes of $\Phi (k)$.
 
We are confident that our result holds (qualitatively) also
for $O(N)$ and $CP(N)$ models. Then for $O(N)$ the success of
the work of Hasenfratz and Niedermayer indicates that the
artifacts are still very small at $N=3$.
 
\section{Conclusions}
 
At $N \to \infty $ the procedure of Hasenfratz and Niedermayer
leads to a perfect scaling in the Gross--Neveu model, as we
showed analytically. In the scaling behavior, cut-off artifacts
are completely eliminated at any correlation length. 
That the FPA really
describes continuum physics on a lattice (of finite spacing)
can also be seen from the fact that the continuum energy spectrum,
and with it continuous Poincar\'{e} invariance, are restored.
 
The small field approximation does not improve the scaling essentially.
Here the desired direction out of the critical surface is projected
on a subspace, which is too restrictive.
 
Also the asymptotic scaling is improved for the FPA, but this is
not of primary importance.

\end{document}